\documentclass[11pt,sort&compress]{elsarticle}

\journal{}
\usepackage{lipsum}
\usepackage{amsfonts}
\usepackage{graphicx}
\usepackage{epstopdf}
\usepackage{algorithmic}
\usepackage{multirow}
\usepackage{algorithm}
\usepackage{algorithmic}

\usepackage{xspace}
\usepackage{epsf}
\usepackage{amsmath}
\usepackage{amssymb}
\usepackage{setspace}
\usepackage{enumerate}
\usepackage{float}
\usepackage{footnote}
\usepackage{scalefnt}
\usepackage{xfrac}
\usepackage{transparent}
\usepackage{rotate}
\usepackage{array}
\usepackage{tabu}
\usepackage[mediumspace,mediumqspace,squaren]{SIunits}
\usepackage{multirow}
\usepackage{enumitem}
\usepackage{tikz}
\usepackage{bm}
\usepackage{microtype}
\usepackage[margin=1in]{geometry}
\usetikzlibrary{patterns.meta}

\usepackage{bm}
\newcommand{\ve}[1]{\bm{#1}}

\newcommand{\ba}{\ve{a}}

\newcommand{\bu}{\ve{u}}

\newcommand{\br}{\ve{r}}

\newcommand{\by}{\ve{y}}
\newcommand{\bx}{\ve{x}}

\newcommand{\bs}{\ve{s}}

\newcommand{\bE}{\ve{E}}

\newcommand{\bP}{\ve{P}}

\newcommand{\dd}{\text{d}}

\newcommand{\eps}{\varepsilon}

\newcommand\Rey{\mbox{\text{Re}}\xspace}

\newcommand{\cD}{\mathcal{D}}
\newcommand{\cI}{\mathcal{I}}

\newcommand{\cL}{\mathcal{L}}
\newcommand{\dL}{\mathbf{L}}
\newcommand{\dW}{\mathbf{W}}

\newcommand{\dM}{\mathbf{M}}
\newcommand{\dU}{\mathbf{U}}
\newcommand{\dD}{\mathbf{D}}

\newcommand{\cbL}{\overline{\mathcal{L}}}

\newcommand{\barc}{\bar{c}}
\newcommand{\bars}{\bar{s}}

\usepackage{hyperref}
\usepackage[]{xcolor}
\usepackage{soul}
\colorlet{soulred}{white}
\sethlcolor{soulred}

\hypersetup{
  colorlinks=true,
}

\definecolor{lightblue}{rgb}{0.63, 0.74, 0.78}
\definecolor{seagreen}{rgb}{0.18, 0.42, 0.41}
\definecolor{orange}{rgb}{0.85, 0.55, 0.13}
\definecolor{silver}{rgb}{0.69, 0.67, 0.66}
\definecolor{rust}{rgb}{0.72, 0.26, 0.06}

\colorlet{lightsilver}{silver!30!white}
\colorlet{darkorange}{orange!75!black}
\colorlet{darksilver}{silver!65!black}
\colorlet{darklightblue}{lightblue!65!black}
\colorlet{darkrust}{rust!85!black}

\usepackage[labelfont=bf,font=small]{caption}

\usepackage{cleveref}
\Crefname{ALC@unique}{Line}{Lines}

\usepackage{comment}
\usepackage{parskip}

\newcommand\blfootnote[1]{%
  \begingroup
  \renewcommand\thefootnote{}\footnote{#1}%
  \addtocounter{footnote}{-1}%
  \endgroup
}
\begin{document}

\hypersetup{
  linkcolor=darkrust,
  citecolor=seagreen,
  urlcolor=darkrust,
  pdfauthor=author,
}

\begin{frontmatter}

\title{{\large\bfseries Fast Macroscopic Forcing Method}}

\author[1]{\vspace{-3ex}Spencer H.\ Bryngelson\corref{cor1}}
\ead{shb@gatech.edu}
\author[1]{Florian Sch\"afer\corref{cor1}}
\ead{florian.schaefer@cc.gatech.edu}
\author[2]{Jessie Liu}
\author[2]{Ali Mani}

\cortext[cor1]{These authors contributed equally to this work (alphabetical by last name).}

\address[1]{School of Computational Science \& Engineering, Georgia Institute of Technology, Atlanta, GA 30332, USA}

\address[2]{Department of Mechanical Engineering, Stanford University, Stanford, California 94305, USA}

\date{}

\begin{abstract}
The macroscopic forcing method (MFM) of \citet{maniMacroscopicForcingMethod2021} and similar methods for obtaining turbulence closure operators, such as the Green's function-based approach of \citet{hamba1995analysis}, recover reduced solution operators from repeated direct numerical simulations (DNS). 
MFM has already been used to successfully quantify Reynolds-averaged Navier--Stokes (RANS)-like operators for homogeneous isotropic turbulence and turbulent channel flows.
Standard algorithms for MFM force each coarse-scale degree of freedom (i.e., degree of freedom in the RANS space) and conduct a corresponding fine-scale simulation (i.e., DNS), which is expensive. 
We combine this method with an approach recently proposed by \citet{schaferSparseRecoveryElliptic2021} to recover elliptic integral operators from a polylogarithmic number of matrix--vector products.
The resulting Fast MFM introduced in this work applies sparse reconstruction to expose local features in the closure operator and reconstructs this coarse-grained differential operator in only a few matrix--vector products and correspondingly, a few MFM simulations. 
For flows with significant nonlocality, the algorithm first ``peels'' long-range effects with dense matrix--vector products to expose a more local operator. 
We demonstrate the algorithm's performance for scalar transport in a laminar channel flow and momentum transport in a turbulent channel flow.
For these problems, we recover eddy diffusivity- and eddy viscosity-like operators, respectively, at $1\%$ of the cost of computing the exact operator via a brute-force approach for the laminar channel flow problem and $13\%$ for the turbulent one.
We observe that we can reconstruct these operators with an increase in accuracy by about a factor of $100$ over randomized low-rank methods. 
Applying these operators to compute the averaged fields of interest has visually indistinguishable behavior from the exact solution. 
Our results show that a similar number of simulations are required to reconstruct the operators to the same accuracy under grid refinement. 
Thus, the accuracy corresponds to the physics of the problem, not the numerics.
We glean that for problems in which the RANS space is reducible to one dimension, eddy diffusivity and eddy viscosity operators can be reconstructed with reasonable accuracy using only a few simulations, regardless of simulation resolution or degrees of freedom. 
\end{abstract}


\end{frontmatter}

\section{Introduction}

\blfootnote{\\ \indent\;\; Code available at: \url{https://github.com/comp-physics/fast-mfm}}

Well-established equations describe even the most complicated flow physics.
Still, full-resolution simulations of them stretch computational resources. 
Reduced-complexity surrogate models are a successful approach to reducing these costs.
Historically, physical insight and analytical techniques have been used to develop these models, including the RANS closure models~\citep{tennekes1972first}.
However, data-driven approaches are emerging as semi-automated tools to accomplish the same task.
Some approaches attempt to represent the time evolution of the physical system via neural networks, a formidable task that involves reducing the entire Navier--Stokes operator~\citep{li2020fourier,lu2021learning}.
An alternative approach is to compute effective equations that act on spatial or temporal averages. 
The governing equations are projected into the reduced or averaged space, and a forcing function is applied to examine the effect of the underlying fluctuations on the averaged behavior. 
~\citet{kraichnan1987eddy} and~\citet{hamba1995analysis} examined Green's function solutions (i.e., using Dirac-delta-function-type forcing) to scalar and momentum transport equations to develop exact expressions for closure operators. 
Similarly, the macroscopic forcing method (MFM) of~\citet{maniMacroscopicForcingMethod2021} quantifies closure operators exactly by examining forcing and averaged responses, called input--output pairs. 
However, as a linear-algebra-based technique, MFM does not require the use Dirac delta functions as forcing basis functions, and others like polynomials~\citep{liu21} and harmonic functions~\citep{shirian22}, can be used. 

MFM has been successfully applied to close reacting flow equations~\citep{shende21,shende22} and analyze homogeneous isotropic turbulence~\citep{shirian22} and turbulent channel flow~\citep{park21}.
MFM is analogous to numerical homogenization, or the finite-dimensional approximation of solution spaces of partial differential equations (PDEs)~\citep{altmann2021numerical}. 
These techniques amount to \textit{operator recovery} or \textit{learning}, where an unknown operator is estimated from a set of input--output pairs obtained from full-resolution simulations.
These simulations are computationally expensive, so there is a pressing need to reduce the number of samples required, which we address in this work.

Using MFM, one constructs effective operators, or \textit{macroscopic} operators, acting on solution averages from full-resolution simulations, called direct numerical simulations (DNS) of the governing or \textit{microscopic} equations~\citep{maniMacroscopicForcingMethod2021}. 
If the macroscopic operators are linear, the MFM procedure is no different from estimating a matrix from a limited number of matrix--vector products.
The number of microscopic simulations required to recover the macroscopic operator exactly equals the number of macroscopic degrees of freedom, which can be prohibitively large for many simulation problems, like high-$\Rey$ turbulence.

One can partially address this problem by working in Fourier space~\citep{maniMacroscopicForcingMethod2021} or fitting a parametric model to approximate the eddy diffusivity operator~\citep{hamba2004nonlocal, park21}. 
However, the former requires spatial homogeneity, and the latter's accuracy depends on the parametric model's quality. 
\citet{liu21} introduces an improved model that uses the nonlocal eddy diffusivity operator's moments to approximate the full operator.
While these are viable approaches and the subject of ongoing work, the target of this work is to reconstruct the full discretely-defined nonlocal eddy diffusivity operator, as opposed to prescribing or modeling its shape. 

For many flows of practical interest, the nonlocal effects of closure terms show diffusive behavior.
Thus, work on operator recovery for elliptic PDEs is closely related to MFM.
\cite{lin2011fast} propose a ``peeling'' approach for recovering hierarchical matrices from a polylogarithmic number of matrix--vector products, although without rigorous bounds on the approximation error.
Extensions of this algorithm were proposed by \cite{martinsson2016compressing,levitt2022randomized,levitt2022linear}.
In this setting, eigendecompositions and randomized linear algebra have been used to recover elliptic solution operators from matrix--vector products~\citep{de2021convergence,boulle2022learning,stepaniants2021learning,nelsen2021random}.
Since the eigenvalues of elliptic solution operators follow a power law, these methods require $\mathrm{poly}(1 / \eps)$ matrix--vector products to obtain an $\eps$ approximation of the operator.
In contrast, \citet{schafer2017compression} showed that hidden sparsity of the solution operator results in an $\eps$ approximation from only $\mathrm{poly}(\log(1/\eps))$ carefully crafted matrix--vector products.
This speedup amounts to an \textit{exponential reduction} in the number of matrix--vector products.
The sparsity used by \citet{schaferSparseRecoveryElliptic2021} results from the locality of the partial differential operator shared by local fluid models.

We use this approach to accelerate the MFM to create the \textit{Fast MFM}. 
The Fast MFM reveals the locality of the physical models to reduce the sample complexity of standard MFM operator recovery.
We apply Fast MFM to inhomogeneous and turbulent problems and reconstruct the RANS closure operators. 
Specifically, we consider passive scalar transport in a laminar 2D channel flow following \citet{maniMacroscopicForcingMethod2021} and reconstruct the corresponding eddy diffusivity operator, and momentum transport in a canonical turbulent 3D channel flow at $\Rey_\tau = 180$ following \citet{hamba2005nonlocal} and \citet{park21} and reconstruct the corresponding eddy viscosity operator.
These examples display sufficient spatio-temporal richness in their dynamics to argue that the Fast MFM can be applied more broadly. 
For example, one could tackle the open closure problems associated with multiphase flows~\citep{bryngelson19,bryngelson19_ML,vie2016particle,ma2016using}, though we do not address such extensions here.

We briefly introduce MFM and similar approaches for recovering turbulence closure operators in \cref{s:mfm}.
\Cref{s:elliptic} details the mathematical foundations of the sparse reconstruction procedure and \cref{s:fmfm} applies it to MFM with an extension to nonsymmetric operators, resulting in the Fast MFM.
Results are presented in \cref{s:results}, focusing on the 2D and 3D problems analyzed by \citet{maniMacroscopicForcingMethod2021} and \citet{park21}.
\Cref{s:conclusions} discusses the outlook of sparse reconstruction methods like the one presented for other flow problems and PDEs broadly.

\section{Background on the Macroscopic Forcing Method (MFM)}\label{s:mfm}

\subsection{The macroscopic forcing method}

Given a set of linear microscopic equations,
\begin{gather}
    \cL c = s,
    \label{e:micro}
\end{gather}
and an averaging operator $(s, c) \mapsto (\bars, \barc)$, the macroscopic (averaged) operator $\cbL$ is defined to satisfy, for all microscopic solutions $c, s$ of \eqref{e:micro}, 
\begin{gather}
    \cbL \barc = \bars.
    \label{e:macro}
\end{gather}
Often, \eqref{e:micro} are advection-diffusion equations for scalar transport or linearizations of nonlinear PDEs such as Navier--Stokes equations, so the averaging operation can be written as 
\begin{gather}
    \bars = \frac{1}{L_2 \cdots L_{N_d}}\int_{\Omega_{2}} \cdots \int_{\Omega_{N_d}} s(x_1, \ldots x_{N_d}) \, \dd x_2 \cdots \dd x_{N_d},
\end{gather}
where 
\begin{gather}
    \Omega = \Omega_{1} \times \Omega_{2} \times \cdots \times \Omega_{N_d}
\end{gather}
is the physical (possibly spatio-temporal) domain and $L_i$ are the lengths in each coordinate direction $x_i$, $i \in \{1,\dots,N_d\}$.
In this example, the averaged \eqref{e:macro} is a univariate problem in the non-averaged coordinate $x_1$. 
However, we point out that the techniques outlined in this work apply to a wider range of possible averaging operations.

\begin{figure}[H]
    \centering
    \includegraphics[scale=1]{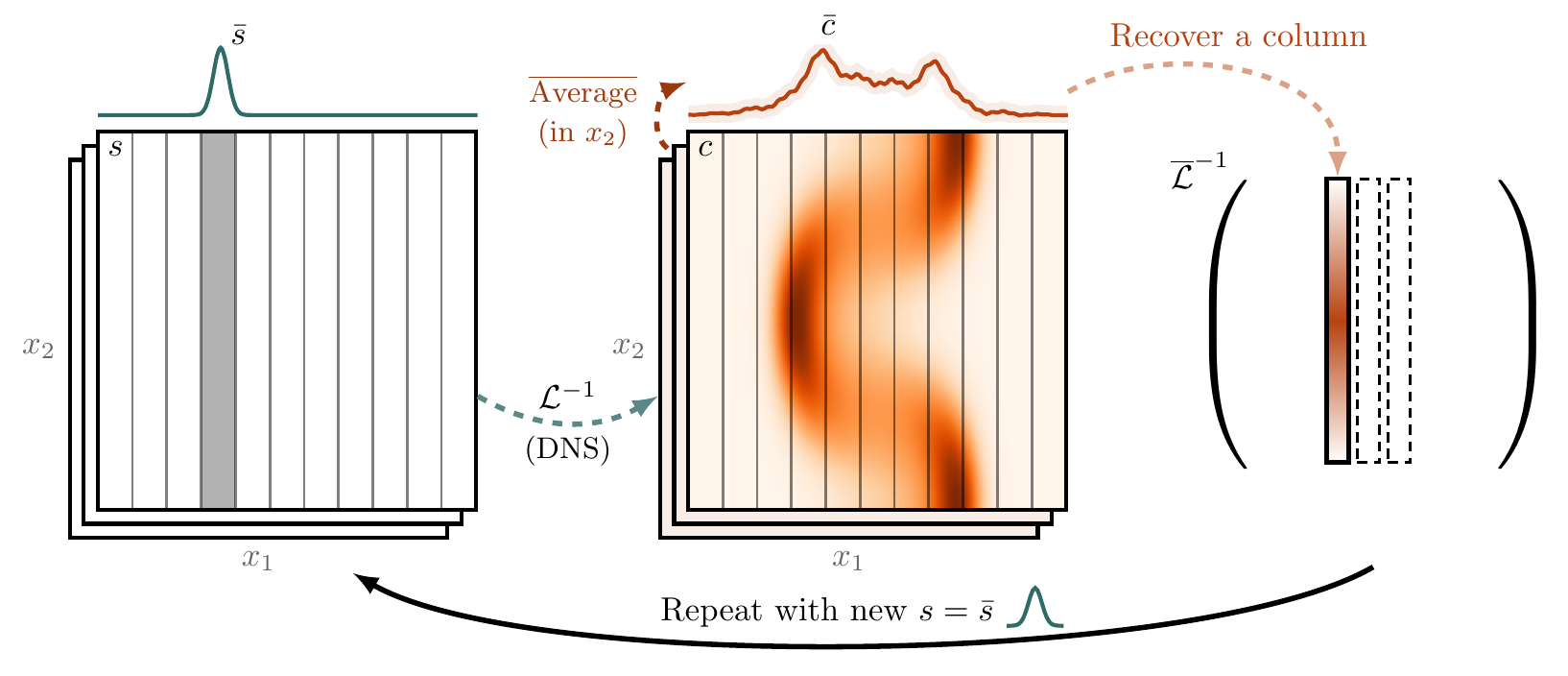}
    \caption{
        Schematic of the MFM. 
    }
    \label{f:overview}
\end{figure}

Using MFM, one can determine the exact linear operator $\cbL$ that acts on averages of flow statistics~\citep{maniMacroscopicForcingMethod2021}.
They infer this operator by generating solution pairs to $\cbL \barc = \bars$, obtained from solving the microscopic equations with forcing $\bars$ and macroscopic solution average $\barc$.

\Cref{f:overview} show an example MFM procedure schematically for a two-dimensional problem with coordinate directions $x_1$ and $x_2$.
The relevant averaging direction is $x_2$, with averaged ``strips'' indicating averaging.
This solution is forced by a field $s(x_1,x_2)$ as a Dirac delta function  at a specific $x_1$ coordinate equivalent to its averaged field $\bars(x_1)$.
The inverse solution operator $\cL^{-1}$ solves the problem \eqref{e:micro} for $c(x_1,x_2)$ given $\bars(x_1)$.
This is computationally equivalent to solving the full-resolution system \eqref{e:micro} or DNS.
The averaged solution field $\barc(x_1)$ corresponds to a column (``recover a column'') of a macroscopic solution operator $\cbL^{-1}$ under this averaging scheme.
This procedure is repeated for all non-averaged degrees of freedom.

In this example, the non-averaged coordinate is $x_1$, so each discretized $x_{1,i}$ is locally forced (via a Dirac delta function) with $s \equiv \bars$.
Completing the MFM procedure gives access to the matrix representation $\cbL$ via $\bs \mapsto \cbL^{-1}\bs$.
Since evaluating this map involves a high-resolution simulation, column-by-column construction of $\cbL^{-1}$ is intractable.

\subsection{The linear algebra of MFM}

A linear algebraic perspective is useful for understanding MFM. 
To this end, we denote as $\cL$ the matrix representation of a discretized advection-diffusion operator 
\begin{gather}
    \label{e:fmm_pde}
    \cL = 
        \frac{\partial}{\partial t}  +
        u \cdot \nabla -
        \nabla \cdot ( a \nabla ),
\end{gather}
where coefficients $a$, $u$, are allowed to vary in space and time. 
The inverse operator, $\cL^{-1}$, takes a spatio-temporal forcing term $s$ as input and returns the spatio-temporal field $c = \cL^{-1} s$ by solving the PDE. 
Let $\bP$ denote a projection onto coarse-scale features of interest, for example, spatio-temporal averages, and $\bE$ denote an extension such that $\bP\bE = \mathcal{I}$, where $\mathcal{I}$ is the identity matrix. 
In the example of \cref{f:overview}, rows of $\bP$  correspond to averages of $c$ in the $x_2$-direction of the domain, and the rows of $\bE$ extend $\bars$ to $s$.
The macroscopic operator can then be expressed as~\citep{maniMacroscopicForcingMethod2021}
\begin{gather}
    \cbL = \left(
        \bP \cL^{-1} \bE \right)^{-1},
\end{gather}
where $\cbL$ and $\cL^{-1}$ are now discretized.

Another perspective on MFM can be obtained considering $\cbL$ in discretized form.
By using bases for its row and column space that consider the row and column spaces of $\bP$ and $\bE$, we obtain a $2 \times 2$ block matrix.
After eliminating the second block, the macroscopic operator is obtained as the Schur complement of $\cL$. 
\begin{gather}
    \cbL
    =
    \left(\left(\cL^{-1}\right)_{1, 1}\right)^{-1}
    =
    \cL_{1, 1} - \cL_{1, 2} \left(\cL_{2,2}\right)^{-1} \cL_{2 ,1}.
    \label{eqn:mfm_schur}
\end{gather}
Computing $\cL^{-1}$ or $\left(\cL_{2, 2}\right)^{-1} \cL_{2, 1}$ naïvely, column by column, requires as many solutions of the full-scale problem as there are coarse scale degrees of freedom. 

\subsection{Inverse MFM}

\begin{figure}
    \centering
    \includegraphics{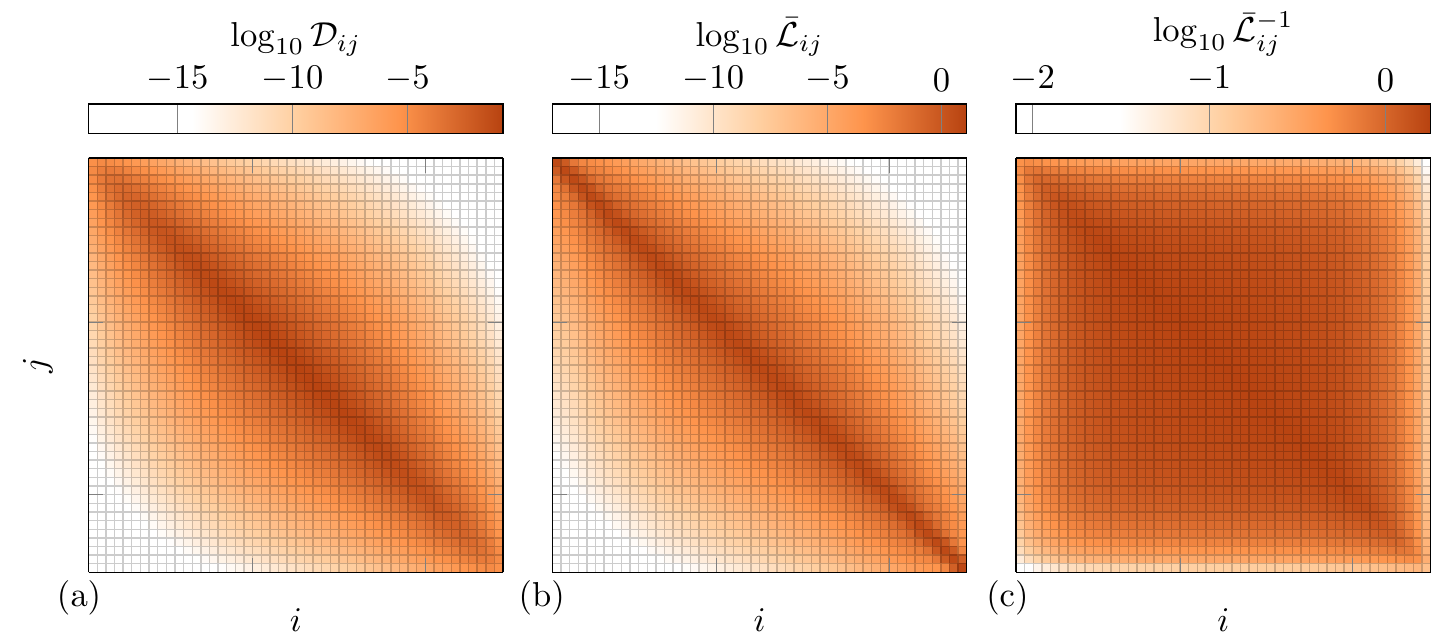}
    \caption{
        Matrices $\cD$, $\cbL$, and $\cbL^{-1}$ as labeled for the 2D channel flow case with 50 non-averaged grid points (and so each is $50 \times 50$) for illustration purposes.
        The inverse operator matrix of (c) is nearly dense, though (a) and (b) are more strongly banded.
        Similar behavior is observed for finer discretizations, which correspond to larger matrices.
    } 
    \label{fig:mfmeddy}
\end{figure}

As shown in \cref{fig:mfmeddy}, $\cbL$ is more local than its computed inverse.
Hoping to turn this locality into computational gains, \citet{maniMacroscopicForcingMethod2021} propose an inverse MFM to directly compute matrix--vector products with $\cbL$ without first having to compute $\cbL^{-1}$ using the ordinary MFM.
This procedure can be interpreted as evaluating the right-hand side of \eqref{eqn:mfm_schur} at the cost of solving a system of equations in $\cL_{2,2}$.
If \eqref{e:micro} is an evolution PDE, this procedure can be interpreted as a control problem, where at each time step, the microscopic portion of the forcing $s$ is chosen to maintain a target average $\barc$. 
The resulting averaged forcing $\bar{s}$ is the same as $\cbL \barc$.

\begin{figure}
    \centering
    \includegraphics{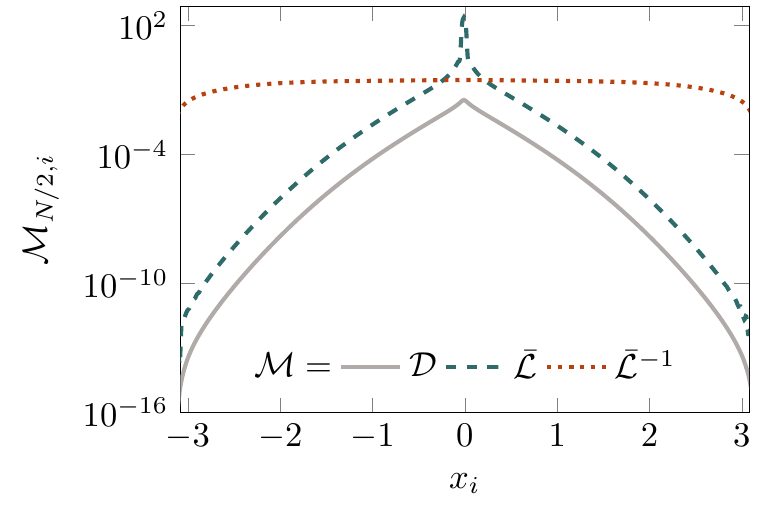}
    \caption{
        The middle row of the matrices of \cref{fig:mfmeddy} on a log scale.
        $\cD$ and $\cbL$ have a similar degree of locality, with entries decreasing in magnitude algebraically from the diagonal.
        The discretized operator $\cbL^{-1}$ is dense; along its diagonal entries only decay modestly at the boundaries.
    } 
    \label{fig:slices}
\end{figure}

\subsection{The eddy diffusivity operator}

As done in \citet{maniMacroscopicForcingMethod2021}, consider a problem in which the averaging operation includes averaging over the entire temporal domain and all directions except $x_1$. 
Using a Reynolds decomposition, the velocity and scalar fields can be decomposed as 
\begin{gather}
    u = \bar{u} + u' \quad \text{and} \quad  c = \bar{c} + c' ,
    \label{e:decomp}
\end{gather}
where $(\cdot)'$ denotes fluctuations about the mean. 
Substitution of \eqref{e:decomp} into the advection--diffusion equation \eqref{e:fmm_pde} with $\cL c = 0$ and averaging results in the corresponding mean scalar equation, we have
\begin{gather}
    \frac{\partial}{\partial x_1} \overline{u_1' c'} - a \frac{\partial^2 \bar{c}}{\partial x_1^2} = 0,
    \label{e:mean_scalar_eqn}
\end{gather}
where the scalar flux, $\overline{u_1'c'}$, is unclosed, and \eqref{e:mean_scalar_eqn} can be written as $\cbL \barc = 0$. 
The macroscopic operator, $\cbL$, contains both the closure for the scalar flux term and the closed molecular diffusion term. 
Thus, $\cbL$ can be further decomposed as
\begin{gather}
   \cbL = - \left(\frac{\partial}{\partial x_1}\right) 
   \left( \cD + a \cI \right) 
   \left( \frac{\partial}{\partial x_1} \right),
   \label{e:eddydiffusivity}
\end{gather}
where $\cD$ is the eddy diffusivity matrix. 
In continuous form, this is equivalent to 
\begin{gather}
    - \overline{ u_1^\prime c^\prime } (x_1) = 
    \int_{y_1} \cD(x_1,y_1)
    \frac{ \partial \barc}{\partial x_1} {\bigg|}_{y_1} \dd y_1,
\end{gather}
which generalizes to the nonlocal eddy diffusivity of \citet{berkowicz1980spectral}:
\begin{gather}
    - \overline{ u_\beta^\prime c^\prime } (\bx) = 
    \int_{\by} \cD^{(\beta\alpha)} (\bx,\by)
    \frac{ \partial \barc}{\partial x_\alpha} {\bigg|}_{\by} \dd \by,
\end{gather}
where $\alpha, \beta \in \{ 1, \dots , N_d \}$ are coordinate directions in the macroscopic space.

As shown in \cref{fig:mfmeddy} and \cref{fig:slices}, the eddy diffusivity matrix is significantly more regular than $\cbL$, making it a preferred target for an operator recovery strategy. 
To leverage these properties in operator recovery, \citet{maniMacroscopicForcingMethod2021} proposed a method for computing matrix--vector products $\cD \ba$ as $\Sigma \cbL \Sigma \ba$, where $\Sigma$ denotes the antiderivative and $\ba$ is the vector state.
The objective of the present work is to recover $\cD$, and thus $\cbL$, from as few matrix--vector products as possible, as accurately as possible.

\section{LU reconstruction of elliptic operators}\label{s:elliptic}

\subsection{Reconstructing elliptic operators from matrix--vector products} 

We use the LU variant of the \emph{Cholesky reconstruction} of \citet{schaferSparseRecoveryElliptic2021} to construct the eddy diffusivity operator.
\citet{schaferSparseRecoveryElliptic2021} prove that the solution operators of divergence form elliptic partial differential equations in dimension $N_d$ can be reconstructed to accuracy $\epsilon$ from only $\mathcal{O}\left(\log^{N_d + 1}(\epsilon^{-1})\right)$ solutions for carefully selected forcing terms. 
We briefly review this approach, which forms the basis of this work.

\subsection{Graph coloring}\label{sec:graph_coloring}

\begin{figure}
    \centering
    \includegraphics{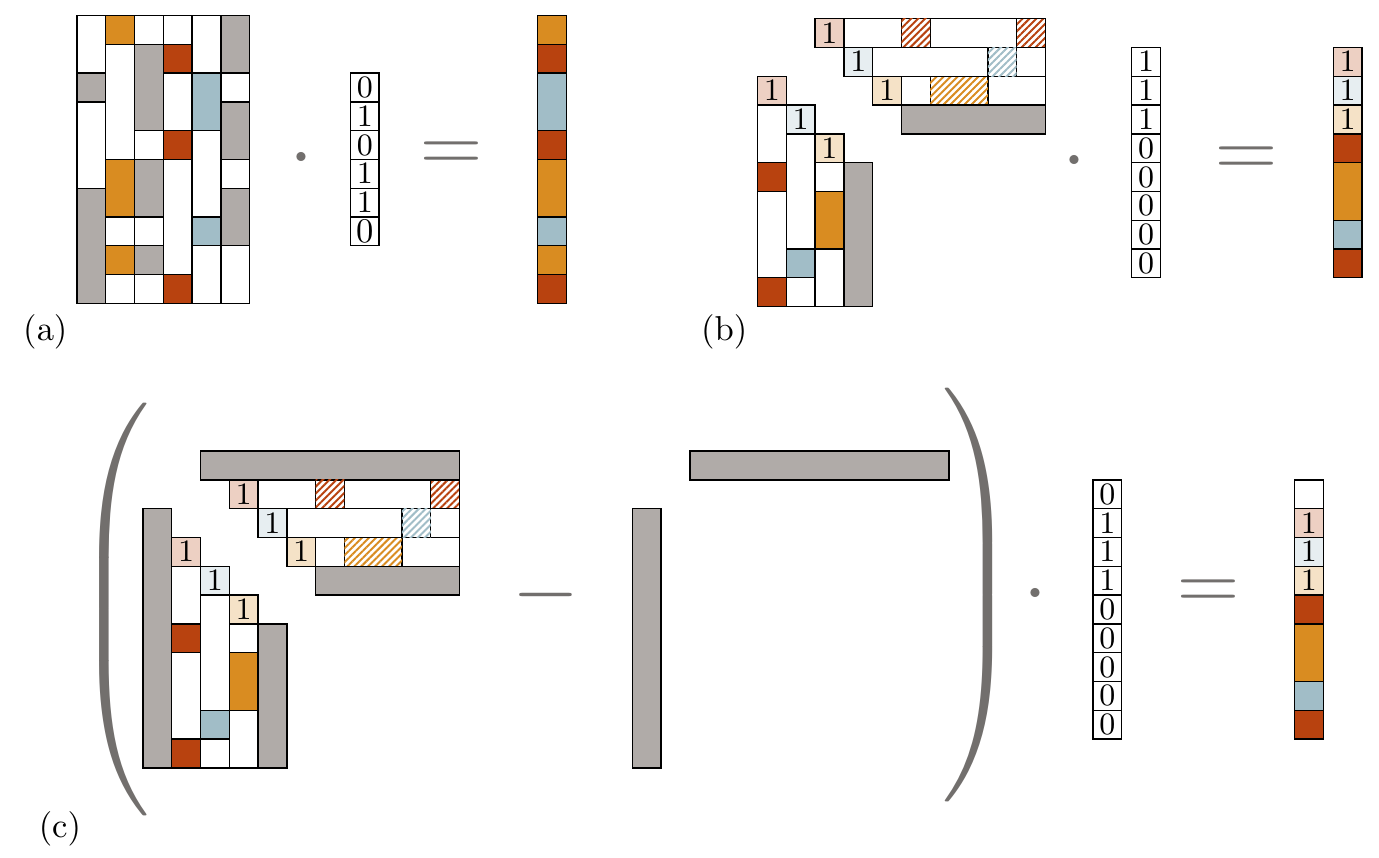}
    \caption{(a) Sparse recovery: Columns with non-overlapping but known sparsity patterns (shown in color) can be recovered by a single matrix--vector product via a carefully chosen vector.
    (b) Factorization: Cholesky factorizations with leading-column sparsity patterns can be recovered similarly. 
    (c) Peeling: If denser columns of the factorization prevent recovery of sparser ones, identify dense columns first and subtract their contribution to recover the sparser ones.
    [This figure was adapted from \citet{schaferSparseRecoveryElliptic2021} with author permission.]
    \label{fig:graph_coloring}}
\end{figure}

\emph{Graph coloring} allows one to reconstruct multiple columns of a sparse matrix from a single matrix--vector product. 
The key idea is to identify groups of columns with non-overlapping sparsity sets and use a right-hand side that only activates those columns.
As illustrated in \cref{fig:graph_coloring}, the selected columns can be read off from the resulting matrix--vector product. 
Similarly, graph coloring can also reveal the leading columns of a sparse LU factorization. 
Once a row-column pair of the LU factors are identified, it can be used to correct the matrix--vector products to reveal later columns. 
This procedure, a variant of which was first proposed by \citet{lin2011fast}, is referred to as peeling (\cref{fig:graph_coloring}).

\subsection{Cholesky factors in wavelet basis}

\begin{figure}
    \centering
    \begin{tabular}{l l l}
        \includegraphics[width=0.33\textwidth]{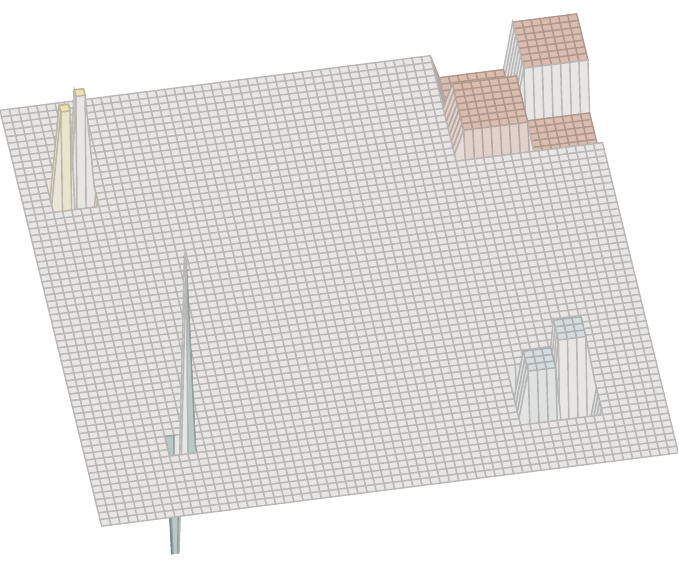} &
        \includegraphics[width=0.27\textwidth]{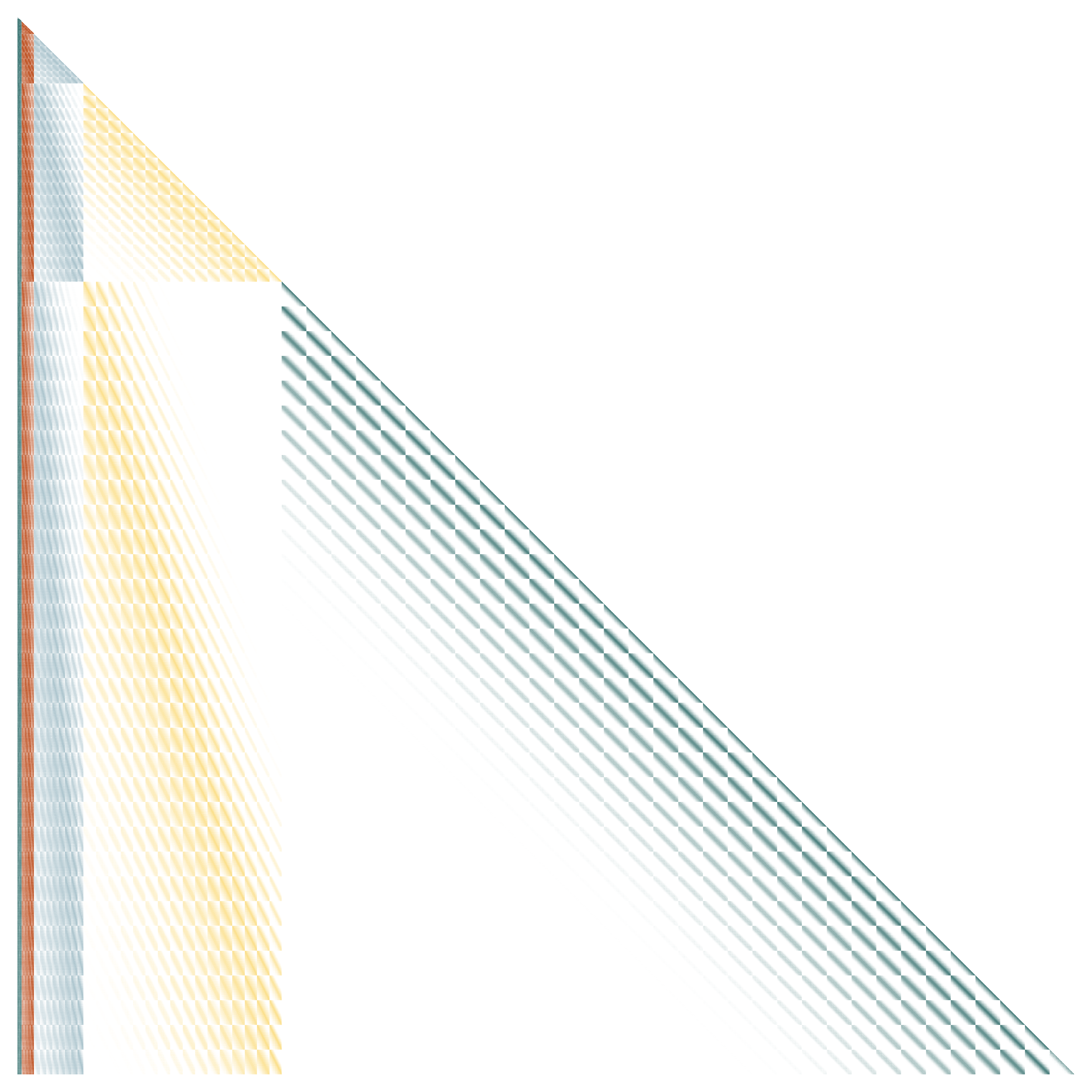} &
        \includegraphics[width=0.33\textwidth]{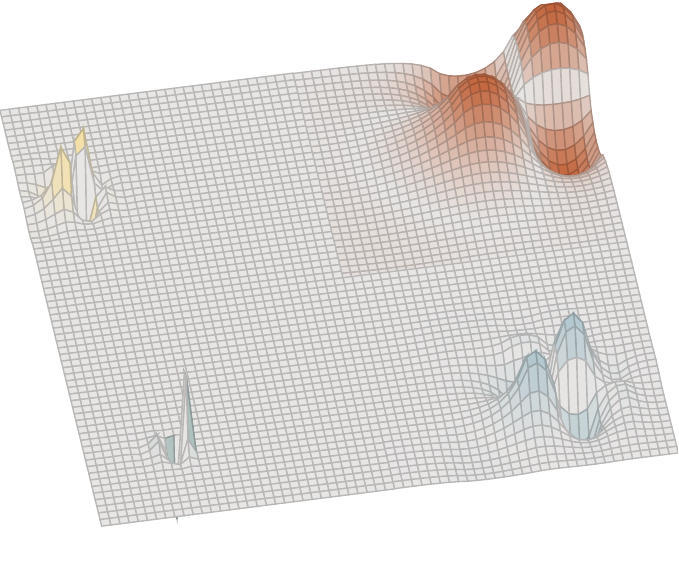}\vspace{-0.5cm} \\
        \tikz\node[fill=white,inner sep=0pt,opacity=1,fill opacity=0.75,text opacity=1]{\small (a)}; & 
        \tikz\node[fill=white,inner sep=0pt,opacity=1,fill opacity=0.75,text opacity=1]{\small (b)}; & 
        \; \tikz\node[fill=white,inner sep=0pt,opacity=1,fill opacity=0.75,text opacity=1]{\small (c)}; 
    \end{tabular}
    \caption{
        (a) shows basis function on four different scales of a multiresolution basis. 
        (b) shows the decay pattern of the Cholesky factorization of an elliptic Green's function discretized in this basis. 
        (c) shows four columns of the Cholesky factor as spatial functions. 
        [This figure was reproduced from \citet{schaferSparseRecoveryElliptic2021} with author permission.]
    }
    \label{fig:choleskydecay}
\end{figure}

It is well-known that the solution operators of elliptic PDEs are dense, owing to the long-range interactions produced by diffusion. 
However, \citet{schafer2017compression} show that when represented in a multiresolution basis ordered \emph{from coarse to fine}, solution operators of elliptic PDEs have almost sparse Cholesky factors. 
This phenomenon is illustrated in \cref{fig:choleskydecay}.
The leading columns of the Cholesky factors, corresponding to a coarse-scale basis function with global support, are dense and therefore limit the efficiency of graph coloring. 
However, they are few and can be identified efficiently and removed via peeling. 
This procedure can be repeated to reveal progressively finer columns. 
The growing number of basis functions on finer scales is compensated for by their smaller support and, thus, increased gains due to graph coloring. 
Thus, the number of matrix--vector products required is approximately constant across levels.
The resulting procedure is described in \cref{alg:lu_recovery}. 
As described in \citet{schaferSparseRecoveryElliptic2021}, the operation $\texttt{scatter}_c$ takes in the vector obtained from a peeled matrix--vector product computed in \cref{line:compute_col,line:compute_row} and uses it to recover the columns associated with the color $c$. 
In principle, further compression of the resulting operator is possible using the techniques of \citet{schafer2021sparse}.

\begin{algorithm}
\begin{algorithmic}[1]
        \setcounter{ALC@unique}{0}
        \STATE $\dL \leftarrow 0 \times N$ empty matrix 
        \STATE $\dU \leftarrow N \times 0$ empty matrix 
        \STATE $\dD \leftarrow 0 \times 0$ empty diagonal matrix 
        \FOR{$c \text{ color}$}
        \label{line:for_loop_cholesky}
        \STATE $\dL_{\mathrm{new}} \leftarrow \texttt{scatter}_c\left( \dW^{\top} \cD \dM_{:, c} - \dL \dD \dU \dW^{\top}\dM_{:, c}\right)$
        \label{line:compute_col}
        \STATE $\dU_{\mathrm{new}} \leftarrow \texttt{scatter}_c\left( \dW^{\top} \cD \dM_{:, c} - \dU^{\top} \dD \dL^{\top} \dW^{\top}\dM_{:, c}\right)$
        \label{line:compute_row}
        \STATE $\dD_{\mathrm{new}} \leftarrow \operatorname{diag}\left(\left(\dL_{\mathrm{new}} + \dU_{\mathrm{new}}\right) / 2\right)^{-1}$
        \STATE $\dL, \dU, \dD \leftarrow 
        \texttt{hcat}\left(\dL, \dL_{\mathrm{new}}\right), 
        \texttt{vcat}\left(\dU, \dU_{\mathrm{new}}\right),
        \texttt{dcat}\left(\dD, \dD_{\mathrm{new}}\right)$ \COMMENT{Concatenate to prior solution}
        \ENDFOR
        \STATE $\dL \leftarrow \dL \dD$
  		\RETURN approximate LU factorization $\dL\dU$ of $\dW^{\top} \cD \dW$
	\end{algorithmic}
\caption{LU reconstruction in wavelet basis given by $\dW$ with measurements given by $\dM$.} \label{alg:lu_recovery}
\end{algorithm}

\subsection{Adaptation to Fast MFM} 

The eddy diffusivity operator is not a divergence-form elliptic solution operator. 
In particular, it is not symmetric. 
Instead of the Cholesky recovery of \citet{schaferSparseRecoveryElliptic2021}, we use an LU recovery that recovers a sparse LU factorization of the target matrix. 
Columns of $\mathrm{L}$ are recovered from matrix--vector products, and rows of $\mathrm{U}$ are recovered from matrix-transpose--vector products (transpose--vector products), which can be computed by solving the adjoint equation of $\cL$.
No rigorous guarantees exist for the accuracy of LU reconstruction applied to eddy diffusivity matrices.
However, \citet{schafer2017compression} show a wide range of diffusion-like operators, including those produced by fractional-order Mat{\'e}rn or Cauchy kernels, produce sparse Cholesky factors, despite the lack of theory supporting this observation.

\subsection{Fast MFM on nonsymmetric operators}\label{s:fmfm}
As remarked by \citet{schaferSparseRecoveryElliptic2021}, a LU version of Cholesky reconstruction that extends to nonsymmetric matrices requires not only matrix--vector products but also transpose--vector products.
Similar requirements arise in hierarchical low-rank approaches~\citep{halko2011finding,lin2011fast}. 
Schur complementation commutes with transposition, in the sense that
\begin{gather}
    \left(\cbL\right)^{\top}
    =
    \left(\left(\cL^{-\top}\right)_{1, 1}\right)^{-1}
    =
    \left(\cL^{\top}\right)_{1, 1} - \left(\cL^{\top}\right)_{1, 2} \left(\left(\cL^\top\right)_{2,2}\right)^{-1} \left(\cL^\top\right)_{2 ,1} .
    \label{eqn:mfm_schur_transpose}
\end{gather}
As a result, transpose--vector products with $\cbL$ can be obtained by applying (inverse or forward) MFM to the $\cbL^{\top}$.
In the case of the discretized advection-diffusion operator in \eqref{e:fmm_pde}, when the system is solved up to time $T$, the transpose of $\cbL$ can be obtained by replacing $v(x, t)$  with $-v(x, T-t)$, $a(x, t)$ with $a(x, T -t)$, and by using the solution at time $T$ as the initial condition. 
Here $x$ is the spatial coordinate, and $t$ denotes time.
The resulting PDE is often called the adjoint problem and frequently arises in the computation of sensitivities of solutions of PDEs with respect to their coefficients, boundary, and initial conditions.
We empirically validate our method using matrix--vector products and transpose--vector products obtained from a full eddy diffusivity operator constructed via brute force (column-by-column) MFM. 
We leave an adjoint-based MFM that efficiently implements transpose--vector products as future work.

\section{Results}\label{s:results}

\subsection{Steady-state laminar channel flow}

Consider a 2D domain representing a channel with left and right walls at $x_1 = \pm \pi$ with Dirichlet boundary condition $c = 0$ and the top and bottom walls with $x_2 = 0,2\pi$ with no flux condition $\partial c/\partial x_2 = 0$.
The scalar field $c(x_1,x_2)$ is governed by a steady advection--diffusion equation with a uniform source term
\begin{gather}
    u_1 \frac{\partial c}{\partial x_1} + u_2 \frac{\partial c}{\partial x_2} = 0.05 \frac{\partial^2 c}{\partial x_1^2} + \frac{\partial^2 c}{\partial x_2^2} + 1,
\end{gather}
where the unequal diffusion constants in the coordinate directions are an outcome of directional nondimensionalization.
The flow is incompressible and satisfies no-penetration boundary conditions on the walls. 
The steady velocity field is prescribed as
\begin{gather}
    u_1 = (1 + \cos(2 x_1))\cos(2 x_2), \quad u_2 =\sin(2 x_1)\sin(2 x_2).
\end{gather}
The PDE is discretized on a uniform staggered mesh with $N_1$ and $N_2 = N_1/2$ grid points in the $x_1$ and $x_2$ coordinate directions. 
Second-order accurate central differences are used. 
The advective fluxes at the cell faces are computed via second-order interpolation and then multiplication of the divergence-free velocity at the face centers.
At the cell faces $x_1=\pm \pi$, the fluxes are computed using ghost points that enforce the specified Dirichlet boundary conditions, while at the cell faces at the top and bottom boundaries, the no flux condition is naturally enforced. 

\begin{figure}
    \centering
    \includegraphics[scale=1]{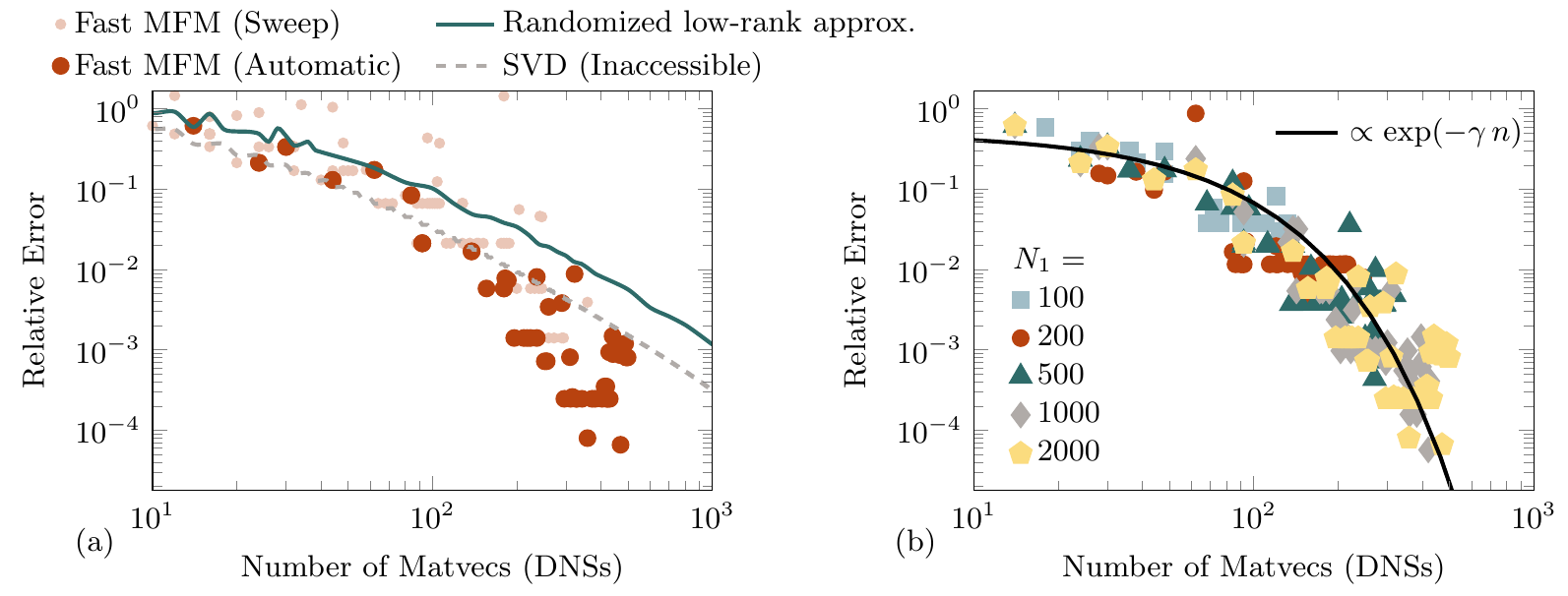}
    \caption{
        Relative $L_2$ errors between the exact MFM operator $\cD$ and its reconstruction for the laminar flow configuration.
        In (b), only the Fast MFM with automatically chosen parameterization is shown, but for different resolutions $N_1$ and the number of matrix--vector products is denoted by $n$ in the legend.
    }
    \label{f:error_peeeling_lam}
\end{figure}

\Cref{f:error_peeeling_lam} shows the macroscopic operator errors for the laminar flow configuration.
In (a), the mesh resolution in the $x_1$ (non-averaged) coordinate is $N_1 = 2000$.
Fast MFM errors are smaller than a truncated SVD reconstruction of the same operator. 
The latter provides the optimal low-rank approximation of $\cD$ but requires access to the full operator and is, therefore, not practical.
A randomized low-rank representation is also shown, which is available.
The errors for this reconstruction are about $10$ times larger than the SVD.
Compared to Fast MFM, these errors are also larger as the number of matrix--vector products increases.
The Fast MFM requires choosing the distance between basis functions of the same color (see \cref{sec:graph_coloring}) and the level at which the wavelet coefficients are truncated. 
The former parameter dictates the cost-accuracy trade-off, and choosing a suitable truncation can improve the method's cost and stability.
A sweep over a wide range of parameters is shown in shaded markers.
We use a heuristic to set these parameters, which results in the non-shaded darker marks. 
Sometimes, the heuristic still produces poor parameter choices resulting in larger Fast MFM errors.

In \cref{f:error_peeeling_lam}~(b), we perform a similar analysis but only show the Fast MFM results for varying mesh sizes $N_1$.
Errors decrease exponentially with the number of matrix--vector products (corresponding to the number of DNSs) with about the same fit coefficients regardless of $N_1$.
This indicates that the Fast MFM reconstruction is dependent on the \textit{physical locality} of the operator, not a numerical or discretized one. 
Thus, operator recovery for high-resolution simulations has an out-sized benefit over traditional MFM. 

\begin{figure}
    \centering
    \includegraphics[scale=1]{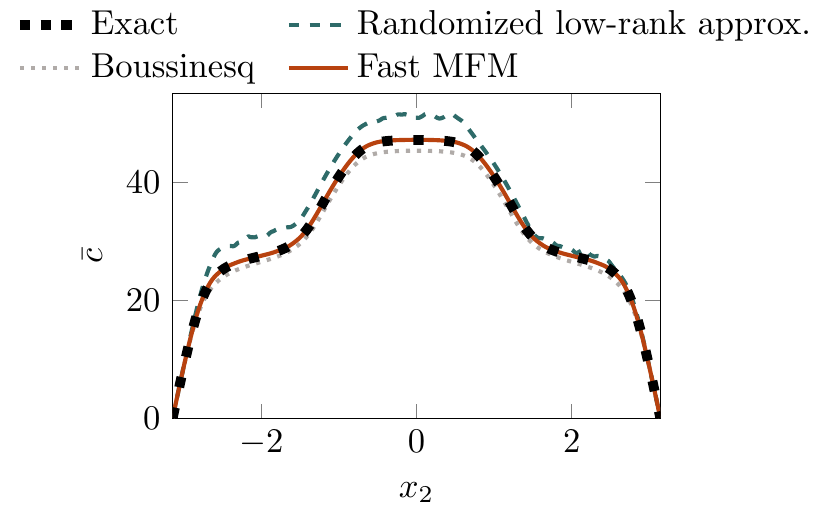}
    \caption{
        Reconstruction of $\barc$ for the laminar, steady channel flow problem. 
        Results are shown for a randomized low-rank approximation, the Boussinesq approximation, and the Fast MFM result. 
        This visualization is 26 out of 2000 possible samples.
    }
    \label{f:lam-reconstructions}
\end{figure}

\Cref{f:lam-reconstructions} shows the application of the recovered operator of \cref{f:error_peeeling_lam} to compute $\barc$ using \eqref{e:eddydiffusivity}. 
The exact result is computed using $N_1 = 2000$ DNSs to recover $\barc$.
For fewer DNSs, $26$ out of the $2000$ total non-averaged degrees of freedom, the Fast MFM result matches the exact result well, but the other methods do not. 
The Boussinesq approximation is purely local \cite{hamba2004nonlocal}:
\begin{gather}
    -\overline{u_1'c'} = D_{\mathrm{Boussinesq}}\frac{\partial \bar{c}}{\partial x_1},
\end{gather}
where
\begin{gather}
    D_{\mathrm{Boussinesq}}(x_1) = 
    \int_{y_1} \cD(x_1,y_1) \dd y_1.
\end{gather}

\subsection{Turbulent channel flow}

We next consider a fully-developed turbulent channel flow, reconstructing eddy diffusivities, which, for momentum transport, are commonly referred to as eddy viscosities.
The incompressible Navier--Stokes equations are
\begin{gather}
    \frac{ \partial u_i}{\partial t} + 
    \frac{\partial u_j u_i}{\partial x_j} = 
    -\frac{\partial p}{\partial x_i} + 
    \frac{1}{\Rey} \frac{\partial^2 u_i}{\partial x_j \partial x_j} + r_i, \\
    \frac{\partial u_j}{\partial x_j} = 0,
\end{gather}
where $\br$ is a body force, $p$ is the pressure, and $\bu$ are velocities.
Following \citet{maniMacroscopicForcingMethod2021}, the generalized momentum transport equation associated with MFM for a computed $u_i$ field is
\begin{gather}
    \frac{ \partial v_i}{\partial t} + 
    \frac{\partial u_j v_i}{\partial x_j} = 
    -\frac{\partial q}{\partial x_i} + 
    \frac{1}{\Rey} \frac{\partial^2 v_i}{\partial x_j \partial x_j} + s_i, \\
    \frac{\partial v_j}{\partial x_j} = 0,
\end{gather}
for a pressure-like term $q$ that ensures incompressibility. 

\begin{figure}
    \centering
    \includegraphics[scale=1]{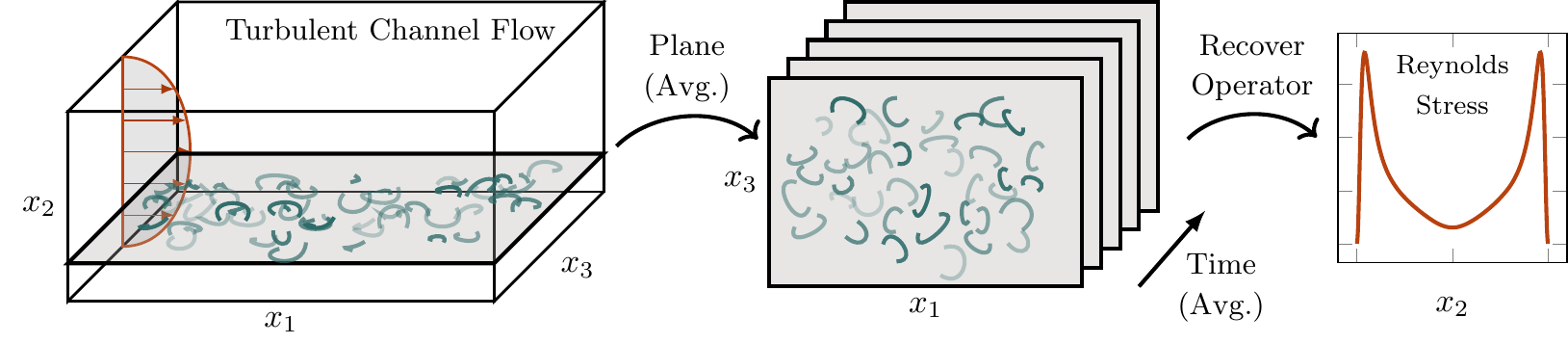}
    \caption{
        Diagram of the Fast MFM reconstruction procedure for the 3D turbulent channel flow case.
    }
    \label{f:schematic-turb}
\end{figure}

We consider a case with $\Rey_\tau = u_\tau \delta / \nu = 180$ where $\delta$ is the channel half-width and $u_\tau$ is the friction velocity.
The mean flow is in the $x_1$ direction, the $x_2$ direction is wall-normal, and the $x_3$ direction is the span with periodic boundaries.
The streamwise domain length is $2\pi$, and the spanwise length is $\pi$.
The body force, $\br$, is the mean pressure gradient in the periodic simulation and is $\br=(1,0,0)$ in this nondimensionalized problem. 
The incompressible Navier--Stokes equations are solved using direct numerical simulation with a $144^3$ grid for $T = 500 \delta/u_\tau$ to ensure statistical convergence.
Simulation-result baselines, including the discretized $\cD$ and $\cL$ matrices, for this case follow from \citet{park21} and are used herein.
\Cref{f:schematic-turb} shows our MFM procedure, averaging all independent variables except for the wall-normal coordinate $x_2$.
We thus recover the Reynolds stresses as a function of $x_2$. The generalized eddy viscosity is given by
\begin{gather}
    - \overline{ u_\beta^\prime u_\gamma^\prime } (\bx) = 
    \int_{\by} \cD^{(\beta\gamma)} (\bx,\by)
    \frac{ \partial \bar{u}_1}{\partial x_2} {\bigg|}_{\by} \dd \by
\end{gather}
where $\cD^{\beta\gamma}$ in our notation represents $\cD_{\beta\gamma21}$ in traditional notation for the eddy viscosity tensor \cite{hamba2005nonlocal}.

\begin{figure}
    \centering
    \includegraphics[scale=1]{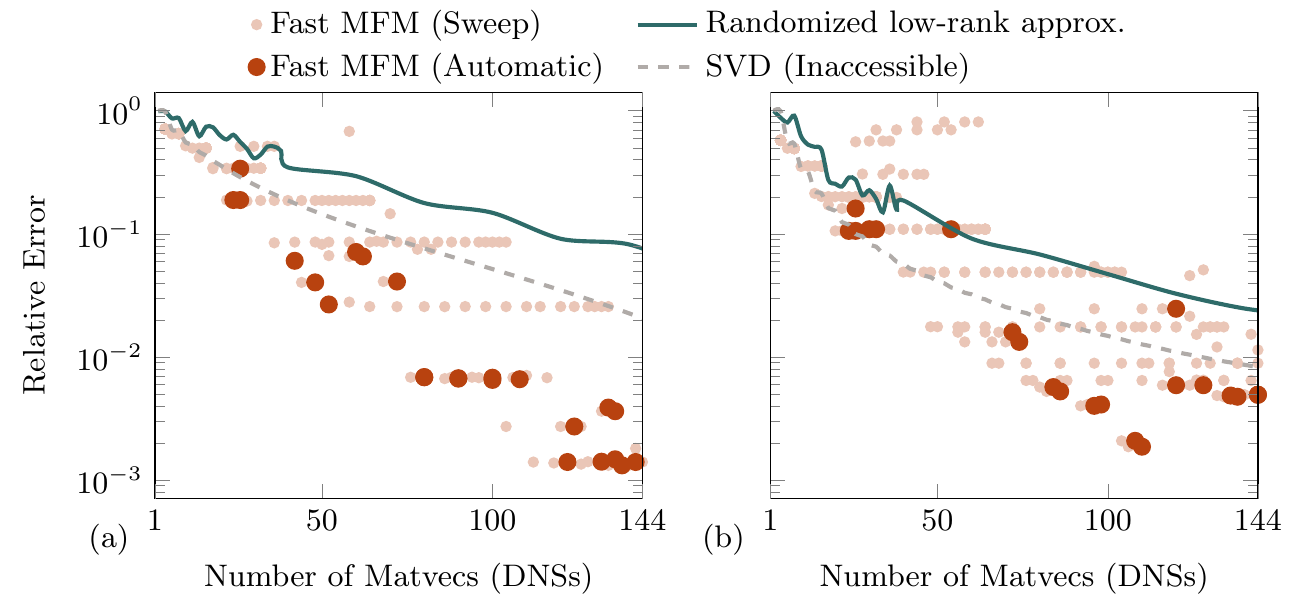}
    \caption{
        Relative errors in the recovered eddy viscosity kernels (a) $\cD^{21}$ and (b) $\cD^{11}$ are shown for the $\Rey_\tau = 180$ turbulent channel flow configuration.
    }
    \label{f:error_peeeling_turb}
\end{figure}

\Cref{f:error_peeeling_turb} shows the errors in the recovered eddy viscosities.
The errors are computed as the difference in operator norms between the approximate and exact solutions to the discretized problem. 
The exact solution is recovered via brute force IMFM, which computes each non-averaged degree of freedom via forcing each $s_i$ independently to recover all columns of $\cD^{\beta\gamma}$  for each $\beta$ and $\gamma$.
In \cref{f:error_peeeling_turb}, the trends of (a) and (b) are similar, with the Fast MFM having smaller errors than both the SVD, which is inaccessible in practice, and a randomized low-rank approximation of it that is accessible.
The differences in errors are small for small numbers of matrix--vector produces as the peeling procedure removes the long-range behaviors.
For larger numbers of matrix--vector products, the difference increases.
Fast MFM has a factor of about 100 smaller errors than the low-rank approximation for 100 matrix--vector products in both (a) and~(b).

\begin{figure}
    \centering
    \includegraphics[scale=1]{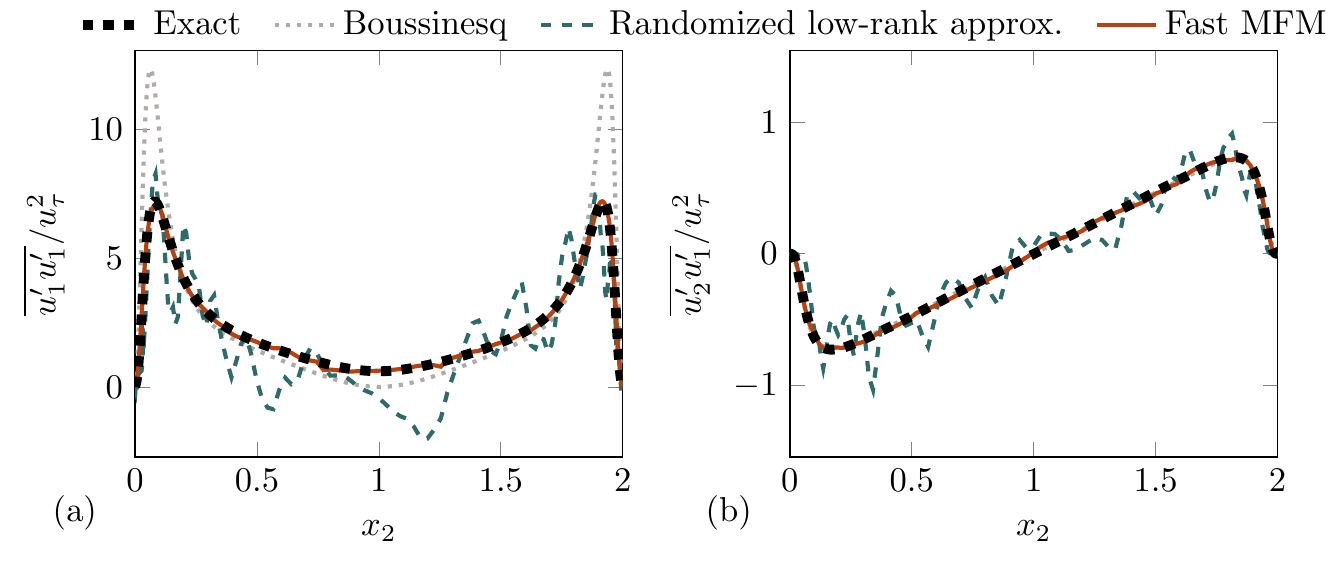}
    \caption{
        Reynolds stress reconstructions (a) $\overline{u_1^\prime u_1^\prime}$ and (b) $\overline{u_2^\prime u_1^\prime}$ for the turbulent channel flow configuration.
        Results are shown for 20 out of 144 possible samples.
    }
    \label{f:reconstructions}
\end{figure}

\Cref{f:reconstructions} shows the Reynolds stress reconstructions for the turbulent channel flow. 
While direct computation of Reynolds stresses needs only one DNS, we use Reynolds stresses as a metric to assess the accuracy of the recovered eddy viscosity operator, which inevitably requires multiple simulations.
The Fast MFM results (solid, thin line) are compared with the Boussinesq approximation and a randomized low-rank procedure.
Exact results are recovered via brute-force MFM.
For 20 simulations, the difference between the exact solution and Fast MFM is not discernible.
For the same number of simulations, the low-rank procedure does not produce a reasonable approximation for either component.
The Boussinesq approximation is a good one for the transverse stress component $\overline{u_2^\prime u_1^\prime}$ of \cref{f:reconstructions}~(b) but does poorly with the $\overline{u_1^\prime u_1^\prime}$ reconstruction in \cref{f:reconstructions}~(a).

\section{Conclusion}\label{s:conclusions}

This work explores a linear algebra approach to reconstructing closure operators.
Fast MFM uses sparse recovery and peeling techniques, revealing local behaviors that crafted forcings can simultaneously recover.
Results show that tens of simulations are required to reconstruct the eddy diffusivity operator and averaged field to visual accuracy.
This contrasts against brute-force MFM, which forces each degree of freedom and is thus prohibitively expensive; the Boussinesq approximation, which is shown to have inaccuracies in some test cases; randomized low-rank approximations, which, while feasible, have poor accuracy; and even SVD, which performs worse than Fast MFM and is inaccessible in a simulation environment.
While other ongoing work focuses on modeling the nonlocal eddy diffusivity using partial differential equations and limited information about the exact eddy diffusivity, the Fast MFM procedure recovers full nonlocal eddy diffusivity operators at low sample complexity. 
It is a stepping stone toward the long-term goal of sample-efficient recovery of coarse-grained time integrators.

\section*{Acknowledgements}

This work used Bridges2 at the Pittsburgh Supercomputing Center through allocation TG-PHY210084 (PI Spencer Bryngelson) from the Advanced Cyberinfrastructure Coordination Ecosystem: Services \& Support (ACCESS) program, which is supported by National Science Foundation grants \#2138259, \#2138286, \#2138307, \#2137603, and \#2138296.
SHB also acknowledges the resources of the Oak Ridge Leadership Computing Facility at the Oak Ridge National Laboratory, which is supported by the Office of Science of the U.S.\ Department of Energy under Contract No.\ DE-AC05-00OR22725. 
SHB acknowledges support from the Office of the Naval Research under grant N00014-22-1-2519 (PM Dr.\ Julie Young).
FS gratefully acknowledges support from the Office of Naval Research under grant N00014-23-1-2545 (PM Dr.\ Reza Malek-Madani).
JL was supported by the Boeing Company. 
AM was supported by the Office of Naval Research under grant N00013-20-1-2718. 
The authors gratefully acknowledge Danah Park for providing the DNS and MFM data of the turbulent channel flow simulations and Dana Lavacot for fruitful discussions.

\bibliographystyle{model1-num-names}
\bibliography{main.bib}

\begin{thebibliography}{31}
\expandafter\ifx\csname natexlab\endcsname\relax\def\natexlab#1{#1}\fi
\providecommand{\url}[1]{\texttt{#1}}
\providecommand{\href}[2]{#2}
\providecommand{\path}[1]{#1}
\providecommand{\DOIprefix}{doi:}
\providecommand{\ArXivprefix}{arXiv:}
\providecommand{\URLprefix}{URL: }
\providecommand{\Pubmedprefix}{pmid:}
\providecommand{\doi}[1]{\href{http://dx.doi.org/#1}{\path{#1}}}
\providecommand{\Pubmed}[1]{\href{pmid:#1}{\path{#1}}}
\providecommand{\bibinfo}[2]{#2}
\ifx\xfnm\relax \def\xfnm[#1]{\unskip,\space#1}\fi
\bibitem[{Mani and Park(2021)}]{maniMacroscopicForcingMethod2021}
\bibinfo{author}{A.~Mani}, \bibinfo{author}{D.~Park},
\newblock \bibinfo{title}{Macroscopic {F}orcing {M}ethod: {A} tool for
  turbulence modeling and analysis of closures},
\newblock \bibinfo{journal}{Physical Review Fluids} \bibinfo{volume}{6}
  (\bibinfo{year}{2021}) \bibinfo{pages}{054607}.
\bibitem[{Hamba(1995)}]{hamba1995analysis}
\bibinfo{author}{F.~Hamba},
\newblock \bibinfo{title}{An analysis of nonlocal scalar transport in the
  convective boundary layer using the green's function},
\newblock \bibinfo{journal}{Journal of Atmospheric Sciences}
  \bibinfo{volume}{52} (\bibinfo{year}{1995}) \bibinfo{pages}{1084--1095}.
\bibitem[{Sch{\"a}fer and Owhadi(2023)}]{schaferSparseRecoveryElliptic2021}
\bibinfo{author}{F.~Sch{\"a}fer}, \bibinfo{author}{H.~Owhadi},
\newblock \bibinfo{title}{Sparse recovery of elliptic solvers from
  matrix--vector products},
\newblock \bibinfo{journal}{arXiv:2110.05351}  (\bibinfo{year}{2023}).
\bibitem[{Tennekes and Lumley(1972)}]{tennekes1972first}
\bibinfo{author}{H.~Tennekes}, \bibinfo{author}{J.~L. Lumley},
  \bibinfo{title}{A first course in turbulence}, \bibinfo{publisher}{MIT
  press}, \bibinfo{year}{1972}.
\bibitem[{Li et~al.(2020)Li, Kovachki, Azizzadenesheli, Bhattacharya, Stuart,
  and Anandkumar}]{li2020fourier}
\bibinfo{author}{Z.~Li}, \bibinfo{author}{N.~B. Kovachki},
  \bibinfo{author}{K.~Azizzadenesheli}, \bibinfo{author}{K.~Bhattacharya},
  \bibinfo{author}{A.~Stuart}, \bibinfo{author}{A.~Anandkumar},
\newblock \bibinfo{title}{Fourier {N}eural {O}perator for parametric partial
  differential equations},
\newblock in: \bibinfo{booktitle}{International Conference on Learning
  Representations}, \bibinfo{year}{2020}, pp. \bibinfo{pages}{1--16}.
\bibitem[{Lu et~al.(2021)Lu, Jin, Pang, Zhang, and
  Karniadakis}]{lu2021learning}
\bibinfo{author}{L.~Lu}, \bibinfo{author}{P.~Jin}, \bibinfo{author}{G.~Pang},
  \bibinfo{author}{Z.~Zhang}, \bibinfo{author}{G.~E. Karniadakis},
\newblock \bibinfo{title}{Learning nonlinear operators via {DeepONet} based on
  the universal approximation theorem of operators},
\newblock \bibinfo{journal}{Nature Machine Intelligence} \bibinfo{volume}{3}
  (\bibinfo{year}{2021}) \bibinfo{pages}{218--229}.
\bibitem[{Kraichnan(1987)}]{kraichnan1987eddy}
\bibinfo{author}{R.~H. Kraichnan},
\newblock \bibinfo{title}{Eddy viscosity and diffusivity: exact formulas and
  approximations},
\newblock \bibinfo{journal}{Complex Systems} \bibinfo{volume}{1}
  (\bibinfo{year}{1987}) \bibinfo{pages}{805--820}.
\bibitem[{Liu et~al.(2021)Liu, Williams, and Mani}]{liu21}
\bibinfo{author}{J.~Liu}, \bibinfo{author}{H.~Williams},
  \bibinfo{author}{A.~Mani},
\newblock \bibinfo{title}{A systematic approach for obtaining and modeling a
  nonlocal eddy diffusivity},
\newblock \bibinfo{journal}{arXiv:2111.03914}  (\bibinfo{year}{2021}).
\bibitem[{Shirian and Mani(2022)}]{shirian22}
\bibinfo{author}{Y.~Shirian}, \bibinfo{author}{A.~Mani},
\newblock \bibinfo{title}{Eddy diffusivity operator in homogeneous isotropic
  turbulence},
\newblock \bibinfo{journal}{Physical Review Fluids} \bibinfo{volume}{7}
  (\bibinfo{year}{2022}) \bibinfo{pages}{L052601}.
\bibitem[{Shende and Mani(2022{\natexlab{a}})}]{shende21}
\bibinfo{author}{O.~B. Shende}, \bibinfo{author}{A.~Mani},
\newblock \bibinfo{title}{Closures for multicomponent reacting flows based on
  dispersion analysis},
\newblock \bibinfo{journal}{Physical Review Fluids} \bibinfo{volume}{7}
  (\bibinfo{year}{2022}{\natexlab{a}}) \bibinfo{pages}{093201}.
\bibitem[{Shende and Mani(2022{\natexlab{b}})}]{shende22}
\bibinfo{author}{O.~B. Shende}, \bibinfo{author}{A.~Mani},
\newblock \bibinfo{title}{A nonlocal extension of dispersion analysis for
  closures in reactive flows},
\newblock \bibinfo{journal}{arXiv:2201.10013}
  (\bibinfo{year}{2022}{\natexlab{b}}).
\bibitem[{Park and Mani(2021)}]{park21}
\bibinfo{author}{D.~Park}, \bibinfo{author}{A.~Mani},
\newblock \bibinfo{title}{Direct calculation of the eddy viscosity operator in
  turbulent channel flow at $\mathrm{Re}_\tau = 180$},
\newblock \bibinfo{journal}{arXiv:2108.10898}  (\bibinfo{year}{2021}).
\bibitem[{Altmann et~al.(2021)Altmann, Henning, and
  Peterseim}]{altmann2021numerical}
\bibinfo{author}{R.~Altmann}, \bibinfo{author}{P.~Henning},
  \bibinfo{author}{D.~Peterseim},
\newblock \bibinfo{title}{Numerical homogenization beyond scale separation},
\newblock \bibinfo{journal}{Acta Numerica} \bibinfo{volume}{30}
  (\bibinfo{year}{2021}) \bibinfo{pages}{1--86}.
\bibitem[{Hamba(2004)}]{hamba2004nonlocal}
\bibinfo{author}{F.~Hamba},
\newblock \bibinfo{title}{Nonlocal expression for scalar flux in turbulent
  shear flow},
\newblock \bibinfo{journal}{Physics of Fluids} \bibinfo{volume}{16}
  (\bibinfo{year}{2004}) \bibinfo{pages}{1493--1508}.
\bibitem[{Lin et~al.(2011)Lin, Lu, and Ying}]{lin2011fast}
\bibinfo{author}{L.~Lin}, \bibinfo{author}{J.~Lu}, \bibinfo{author}{L.~Ying},
\newblock \bibinfo{title}{Fast construction of hierarchical matrix
  representation from matrix--vector multiplication},
\newblock \bibinfo{journal}{Journal of Computational Physics}
  \bibinfo{volume}{230} (\bibinfo{year}{2011}) \bibinfo{pages}{4071--4087}.
\bibitem[{Martinsson(2016)}]{martinsson2016compressing}
\bibinfo{author}{P.-G. Martinsson},
\newblock \bibinfo{title}{Compressing rank-structured matrices via randomized
  sampling},
\newblock \bibinfo{journal}{SIAM Journal on Scientific Computing}
  \bibinfo{volume}{38} (\bibinfo{year}{2016}) \bibinfo{pages}{A1959--A1986}.
\bibitem[{Levitt and Martinsson(2022{\natexlab{a}})}]{levitt2022randomized}
\bibinfo{author}{J.~Levitt}, \bibinfo{author}{P.-G. Martinsson},
\newblock \bibinfo{title}{Randomized compression of rank-structured matrices
  accelerated with graph coloring},
\newblock \bibinfo{journal}{arXiv:2205.03406}
  (\bibinfo{year}{2022}{\natexlab{a}}).
\bibitem[{Levitt and Martinsson(2022{\natexlab{b}})}]{levitt2022linear}
\bibinfo{author}{J.~Levitt}, \bibinfo{author}{P.-G. Martinsson},
\newblock \bibinfo{title}{Linear-complexity black-box randomized compression of
  hierarchically block separable matrices},
\newblock \bibinfo{journal}{arXiv:2205.02990}
  (\bibinfo{year}{2022}{\natexlab{b}}).
\bibitem[{{de Hoop} et~al.(2023){de Hoop}, Kovachki, Nelsen, and
  Stuart}]{de2021convergence}
\bibinfo{author}{M.~V. {de Hoop}}, \bibinfo{author}{N.~B. Kovachki},
  \bibinfo{author}{N.~H. Nelsen}, \bibinfo{author}{A.~M. Stuart},
\newblock \bibinfo{title}{Convergence rates for learning linear operators from
  noisy data},
\newblock \bibinfo{journal}{SIAM/ASA Journal on Uncertainty Quantification}
  \bibinfo{volume}{11} (\bibinfo{year}{2023}) \bibinfo{pages}{480--513}.
\bibitem[{Boull{\'e} and Townsend(2022)}]{boulle2022learning}
\bibinfo{author}{N.~Boull{\'e}}, \bibinfo{author}{A.~Townsend},
\newblock \bibinfo{title}{Learning elliptic partial differential equations with
  randomized linear algebra},
\newblock \bibinfo{journal}{Foundations of Computational Mathematics}
  (\bibinfo{year}{2022}) \bibinfo{pages}{1--31}.
\bibitem[{Stepaniants(2021)}]{stepaniants2021learning}
\bibinfo{author}{G.~Stepaniants},
\newblock \bibinfo{title}{Learning partial differential equations in
  reproducing kernel {H}ilbert spaces},
\newblock \bibinfo{journal}{arXiv:2108.11580}  (\bibinfo{year}{2021}).
\bibitem[{Nelsen and Stuart(2021)}]{nelsen2021random}
\bibinfo{author}{N.~H. Nelsen}, \bibinfo{author}{A.~M. Stuart},
\newblock \bibinfo{title}{The random feature model for input-output maps
  between {B}anach spaces},
\newblock \bibinfo{journal}{SIAM Journal on Scientific Computing}
  \bibinfo{volume}{43} (\bibinfo{year}{2021}) \bibinfo{pages}{A3212--A3243}.
\bibitem[{Sch\"{a}fer et~al.(2021)Sch\"{a}fer, Sullivan, and
  Owhadi}]{schafer2017compression}
\bibinfo{author}{F.~Sch\"{a}fer}, \bibinfo{author}{T.~J. Sullivan},
  \bibinfo{author}{H.~Owhadi},
\newblock \bibinfo{title}{Compression, inversion, and approximate {PCA} of
  dense kernel matrices at near-linear computational complexity},
\newblock \bibinfo{journal}{SIAM Multiscale Modeling \& Simulation}
  \bibinfo{volume}{19} (\bibinfo{year}{2021}) \bibinfo{pages}{688--730}.
\bibitem[{Hamba(2005)}]{hamba2005nonlocal}
\bibinfo{author}{F.~Hamba},
\newblock \bibinfo{title}{Nonlocal analysis of the {R}eynolds stress in
  turbulent shear flow},
\newblock \bibinfo{journal}{Physics of Fluids} \bibinfo{volume}{17}
  (\bibinfo{year}{2005}) \bibinfo{pages}{115102}.
\bibitem[{Bryngelson et~al.(2019)Bryngelson, Schmidmayer, and
  Colonius}]{bryngelson19}
\bibinfo{author}{S.~H. Bryngelson}, \bibinfo{author}{K.~Schmidmayer},
  \bibinfo{author}{T.~Colonius},
\newblock \bibinfo{title}{A quantitative comparison of phase-averaged models
  for bubbly, cavitating flows},
\newblock \bibinfo{journal}{International Journal of Multphase Flow}
  \bibinfo{volume}{115} (\bibinfo{year}{2019}) \bibinfo{pages}{137--143}.
\bibitem[{Bryngelson et~al.(2020)Bryngelson, Charalampopoulos, Sapsis, and
  Colonius}]{bryngelson19_ML}
\bibinfo{author}{S.~H. Bryngelson}, \bibinfo{author}{A.~Charalampopoulos},
  \bibinfo{author}{T.~P. Sapsis}, \bibinfo{author}{T.~Colonius},
\newblock \bibinfo{title}{A {G}aussian moment method and its augmentation via
  {LSTM} recurrent neural networks for the statistics of cavitating bubble
  populations},
\newblock \bibinfo{journal}{International Journal of Multiphase Flow}
  \bibinfo{volume}{127} (\bibinfo{year}{2020}) \bibinfo{pages}{103262}.
\bibitem[{Vi{\'e} et~al.(2016)Vi{\'e}, Pouransari, Zamansky, and
  Mani}]{vie2016particle}
\bibinfo{author}{A.~Vi{\'e}}, \bibinfo{author}{H.~Pouransari},
  \bibinfo{author}{R.~Zamansky}, \bibinfo{author}{A.~Mani},
\newblock \bibinfo{title}{Particle-laden flows forced by the disperse phase:
  {C}omparison between {L}agrangian and {E}ulerian simulations},
\newblock \bibinfo{journal}{International Journal of Multiphase Flow}
  \bibinfo{volume}{79} (\bibinfo{year}{2016}) \bibinfo{pages}{144--158}.
\bibitem[{Ma et~al.(2016)Ma, Lu, and Tryggvason}]{ma2016using}
\bibinfo{author}{M.~Ma}, \bibinfo{author}{J.~Lu},
  \bibinfo{author}{G.~Tryggvason},
\newblock \bibinfo{title}{Using statistical learning to close two-fluid
  multiphase flow equations for bubbly flows in vertical channels},
\newblock \bibinfo{journal}{International Journal of Multiphase Flow}
  \bibinfo{volume}{85} (\bibinfo{year}{2016}) \bibinfo{pages}{336--347}.
\bibitem[{Berkowicz and Prahm(1980)}]{berkowicz1980spectral}
\bibinfo{author}{R.~Berkowicz}, \bibinfo{author}{L.~P. Prahm},
\newblock \bibinfo{title}{On the spectral turbulent diffusivity theory for
  homogeneous turbulence},
\newblock \bibinfo{journal}{Journal of Fluid Mechanics} \bibinfo{volume}{100}
  (\bibinfo{year}{1980}) \bibinfo{pages}{433--448}.
\bibitem[{Sch\"{a}fer et~al.(2021)Sch\"{a}fer, Katzfuss, and
  Owhadi}]{schafer2021sparse}
\bibinfo{author}{F.~Sch\"{a}fer}, \bibinfo{author}{M.~Katzfuss},
  \bibinfo{author}{H.~Owhadi},
\newblock \bibinfo{title}{Sparse {C}holesky factorization by
  {K}ullback--{L}eibler minimization},
\newblock \bibinfo{journal}{SIAM Journal on Scientific Computing}
  \bibinfo{volume}{43} (\bibinfo{year}{2021}) \bibinfo{pages}{A2019--A2046}.
\bibitem[{Halko et~al.(2011)Halko, Martinsson, and Tropp}]{halko2011finding}
\bibinfo{author}{N.~Halko}, \bibinfo{author}{P.-G. Martinsson},
  \bibinfo{author}{J.~A. Tropp},
\newblock \bibinfo{title}{Finding structure with randomness: Probabilistic
  algorithms for constructing approximate matrix decompositions},
\newblock \bibinfo{journal}{SIAM Review} \bibinfo{volume}{53}
  (\bibinfo{year}{2011}) \bibinfo{pages}{217--288}.

\end{thebibliography}

\end{document}